\documentclass[conference]{IEEEtran}
\IEEEoverridecommandlockouts
\usepackage{cite}
\usepackage{amsmath,amssymb,amsfonts}
\usepackage{algorithmic}
\usepackage{graphicx}
\usepackage{textcomp}
\usepackage{xcolor}
\usepackage{hyperref}
\usepackage{url}
\usepackage{tabularx}
\usepackage[table]{xcolor} % 用于表格颜色填充
\usepackage{float}
\usepackage{booktabs}
\usepackage{multirow}
\usepackage{cleveref}
\usepackage{mathrsfs} % 加载 mathrsfs 宏包
\usepackage{graphicx} % 用于插入图片
\usepackage{subcaption} % 用于支持子图
\usepackage{makecell}
\usepackage{colortbl}       % 支持颜色填充
\usepackage{array}          % 调整列间距
\usepackage{caption}        % 控制表格标题格式
\def\BibTeX{{\rm B\kern-.05em{\sc i\kern-.025em b}\kern-.08em
    T\kern-.1667em\lower.7ex\hbox{E}\kern-.125emX}}
\definecolor{headergray}{RGB}{235,235,235}
\definecolor{lightgreen}{RGB}{217, 245, 217}
\definecolor{headerblue}{RGB}{210,230,255}
\definecolor{lightgreenn}{RGB}{220, 250, 220}
\usepackage{listings}
\begin{document}

\title{AssertMiner: Module-Level Spec Generation and Assertion Mining using Static Analysis Guided LLMs
\thanks{\textdagger~ Corresponding Author
\\This paper has been published in the 31st Asia and South Pacific Design Automation Conference (ASP-DAC 2026), Jan. 19-22, 2026, Hong Kong Disneyland.}
}

%A Static Analysis–Guided LLM Framework for Fine-Grained Assertion Generation in Module-Level Verification

\author{
\IEEEauthorblockN{
Hongqin Lyu\textsuperscript{1,2},
Yonghao Wang\textsuperscript{1},
Jiaxin Zhou\textsuperscript{3}, 
Zhiteng Chao\textsuperscript{1\textdagger},
Tiancheng Wang\textsuperscript{1},
and
Huawei Li\textsuperscript{1,2\textdagger}}
\IEEEauthorblockA{\textsuperscript{1}State Key Lab of Processors, Institute of Computing Technology, CAS, Beijing, China}
\IEEEauthorblockA{\textsuperscript{2}University of Chinese Academy of Sciences, Beijing, China}
\IEEEauthorblockA{\textsuperscript{3}Beijing Normal University}
\IEEEauthorblockA{\{lvhongqin24b, wangyonghao22s, chaozhiteng, wangtiancheng, lihuawei\}@ict.ac.cn}
\IEEEauthorblockA{202321130074@mail.bnu.edu.cn}
}

\maketitle

\begin{abstract}
Assertion-based verification (ABV) is a key approach to checking whether a logic design complies with its architectural specifications. Existing assertion generation methods based on design specifications typically produce only top-level assertions, overlooking verification needs on the implementation details in the modules at the micro-architectural level, where design errors occur more frequently. To address this limitation, we present AssertMiner, a module-level assertion generation framework that leverages static information generated from abstract syntax tree (AST) to assist LLMs in mining assertions. Specifically, it performs AST-based structural extraction to derive the module call graph, I/O table, and dataflow graph, guiding the LLM to generate module-level specifications and mine module-level assertions. Our evaluation demonstrates that AssertMiner outperforms existing methods such as AssertLLM and Spec2Assertion in generating high-quality assertions for modules. When integrated with these methods, AssertMiner can enhance the structural coverage and significantly improve the error detection capability, enabling a more comprehensive and efficient verification process.
\end{abstract}


\section{Introduction}
Functional verification is essential in the design process of integrated circuits (ICs), where verification engineers check whether the register-transfer level (RTL) code meets the architectural specifications. Assertion-based verification (ABV) is widely adopted in RTL designs due to its ability to enhance observability and reduce simulation debugging time by up to 50\% \cite{b1, b2}. In particular, high-quality SystemVerilog assertions (SVA) for formal property verification (FPV) are critical within ABV methodologies \cite{b3,b4}, particularly when they could accurately reflect both high-level design intent and low-level RTL details. However, specifications often contain ambiguities \cite{b5,b6}, and different designers may implement the same specification in significantly different ways, which makes translating specifications into effective SVA assertions a time-consuming and labor-intensive process \cite{b7}. As design complexity increases, developing more efficient methods for SVA generation has become an urgent necessity.

\begin{figure}[h]
\centering
\includegraphics[width=0.9\linewidth]{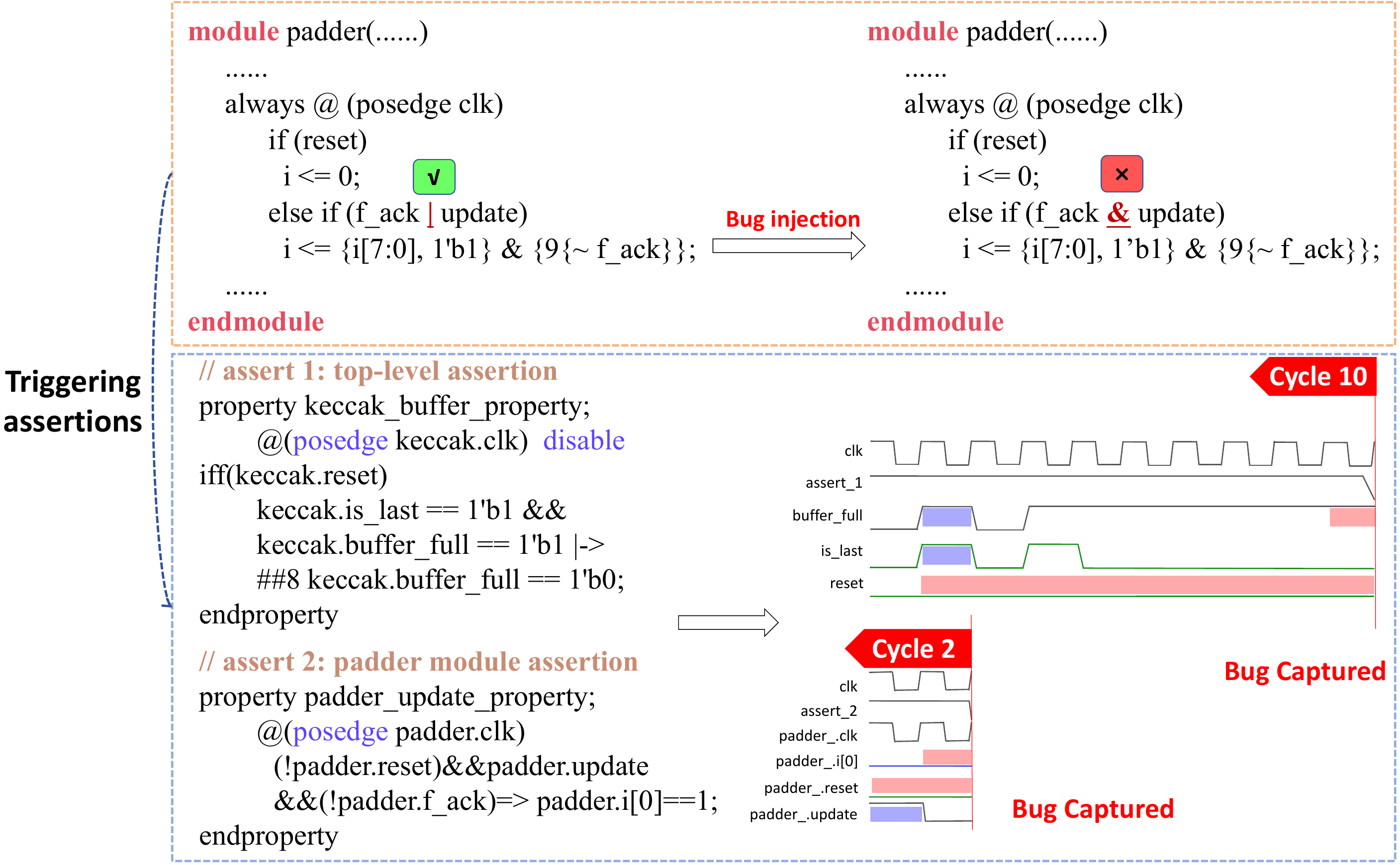}
\caption{A case of rapid assertion triggering and efficient debugging via assertion in the padder module of SHA3 \cite{b28}.}
\label{fig:1}
\end{figure}

The application of large language models (LLMs) in the hardware domain has inspired new approaches to generating assertions for hardware design. Research on automatically generating SVA using LLMs can be broadly divided into two categories. The first category generates assertions using both the specification and RTL code \cite{b8, b9, b10},  which helps capture more detailed functional characteristics. However, RTL-derived assertions may inherit RTL bugs that conflict with the specification, weakening their ability to catch internal errors. The second category relies solely on LLMs to translate a natural language specification into assertions without using RTL code\cite{b11, b12, b13, b14, b15}. These methods are tend to miss assertions that verify implementation details present in RTL. To alleviate this problem, some studies have sought to improve assertion-generation accuracy by establishing mappings between the specification and assertions \cite{b16}.

More importantly, existing assertion generation methods mainly focus on extracting high-level functional logic from specifications, while neglecting the necessity of assertions targeting low-level modules. In modern SoCs, many bugs lie in internal logic chains beyond the reach of top-level assertions. It is either difficult to capture them with top-level assertions, or challenging to pinpoint the exact module or signal path where they occur. In contrast, deep assertions, namely assertions embedded deeply into the low-level modules, can monitor properties at a finer granularity within the module hierarchy, helping to detect errors deeply embedded in the module hierarchy and locate error sources more quickly, thereby significantly reducing debugging time.

Fig. \ref{fig:1} gives a case study to demonstrate the advantage of module-level deep assertions in detecting and localizing design errors over top-level assertions. In this case study, a single error (which mistakenly changes the operator ``$|$'' to ``$\&$'') is injected into a module named padder. The error can trigger both an assertion in the top-level design and a deep assertion within the module simultaneously, as shown in Fig. \ref{fig:1}. It takes ten clock cycles to trigger the top-level assertion while the error’s location needs further analysis. In contrast, it takes only two clock cycles to trigger the module-level deep assertion and the error signal is directly identified, significantly reducing the debugging time.
 
% The key challenge in addressing this problem lies in accurately extracting the specification of each module from the existing raw specifications and RTL code. This is not trivial, as the RTL implementation itself may contain design errors, making it an unreliable source for specification inference. In practical verification, module input and output port configurations are typically assumed to be correct; otherwise, the verification process would lose its validity. Therefore, by analyzing these port configurations together with the inter-module invocation relationships defined at the top level, the overall functionality and signal logic of each module can be inferred. Given the semantic understanding and contextual reasoning capabilities of LLMs, such inference is indeed feasible \cite{16}. %这里的这个描述还是需要进行修改的

Motivated by the importance of deep assertions, this paper proposes AssertMiner, a framework designed to automate deep assertion mining for hierarchical hardware designs. AssertMiner employs abstract syntax tree (AST) to extract static structure information from RTL code. It then leverages the semantic understanding capabilities of LLMs to infer module-level specifications from the static structure information, without relying on RTL implementation details (thus avoiding contamination by potential design errors inside modules). The generated module-level specifications further enables LLMs for effective deep assertion mining.

% This paper presents a deep assertion generation framework named AssertMiner. The framework employs static analysis to automatically extract module port information and inter-module invocation relationships. It then leverages the semantic understanding capabilities of LLMs to infer module-level specifications without relying on RTL implementation details, thereby facilitating the generation of deep assertions.

The contributions of this paper are summarized as follows:

\begin{enumerate}[]
    \item We figure out that the missing of module-level specifications becomes the main obstacle to the generation of effective deep assertions using LLMs. For the purpose of module-level spec generation without relying on RTL implementation details, we propose to use AST to extract static structural information from RTL code, which guides an LLM to generate accurate and consistent module-level specifications.
    \item The proposed AssertMiner integrates multiple LLMs for module-level spec generation, verification feature extraction, and assertion mining, enhancing the capability to capture design errors inside low-level modules.
    \item Experimental results show that the generated module-level assertions by AssertMiner exhibit high structural coverage, and when integrated with the SOTA LLM-based assertion generation methods, can significantly improve the error detection capability, enabling a more comprehensive and efficient verification process.
\end{enumerate}

\section{Related work}
\subsection{Assertion Generation Based on NLP}
Schema-based assertion generation\cite{b17} is one of the early attempts at automatic assertion generation through the static analysis of natural language processing (NLP). It constructs procedural models by instantiating templates with design constructs and constraints. 

However, this method mainly targets combinational logic or static properties and cannot generate assertions with temporal characteristics. Therefore, subsequent studies \cite{b18,b19,b20,b21} further explored automatic generation mechanisms for temporal assertions based on this approach. Static and dynamic analysis techniques were introduced to capture temporal dependencies among design signals, thereby enabling the automated derivation of sequential or time-constrained assertions. Meanwhile, to prevent RTL design flaws from propagating and influencing the generated assertions, \cite{b22} proposed to employ subtree analysis to generate assertions directly from the specification. Building on this foundation, Ganapathy P. et al.\cite{b23} employed sentence classification, named entity recognition, and enhanced parse tree analysis to automatically generate assertions, thereby reducing the workload of writing assertions while simultaneously lowering error rates. However, purely machine-learning methods are susceptible to issues such as long sentences and grammatical errors, leading to inaccuracies in the generated output. Consequently, Iman et al.\cite{b24} proposed a hybrid rule-based and machine-learning approach to improve the accuracy of assertion generation.

 Due to the lack of semantic relevance understanding in traditional NLP methods, leveraging LLMs with stronger semantic comprehension has emerged as a promising direction for improvement in assertion generation.

\subsection{Assertion Generation Based on LLMs}

% The development of automated hardware assertion generation frameworks has advanced significantly, leveraging the robust semantic understanding of LLMs. Recent studies have focused on refining these frameworks to improve efficiency and accuracy. Rahul Kande et al. \cite{12} were early explorers of using LLMs for hardware security assertions. Assertllm by Fang et al. \cite{10} was the first method to process complete specification files and generate comprehensive SVAs for each architectural signal. Bai et al. \cite{6} introduced AssertionForge, which constructed a unified Knowledge Graph (KG) from both natural language specifications and RTL code, bridging the gap between high-level intent and low-level details to enhance assertion quality. Wu et al. \cite{14} proposed Spec2Assertion, which used progressive regularization and Chain-of-Thought prompting to generate high-quality assertions directly from specifications, achieving superior syntactic correctness and functional relevance.

Building upon their acknowledged natural language processing capabilities, a growing body of research has begun to explore LLMs in hardware assertion generation. Rahul Kande et al.\cite{b13} were the first to investigate the feasibility of leveraging LLMs to automatically generate hardware security assertions. To explore the adaptability of LLMs in the domain of Assertion-Based Verification (ABV), Maddala K et al.\cite{b25} proposed an LLM-assisted assertion generation framework tailored for dynamic simulation scenarios. Meanwhile, Mali B. et al.\cite{b26} leveraged GPT-4 to standardize and structure design specifications, and incorporated a simulation log feedback loop to enable automated assertion generation and iterative refinement.

\begin{figure*}[h]
\centering
\includegraphics[width=0.92\linewidth]{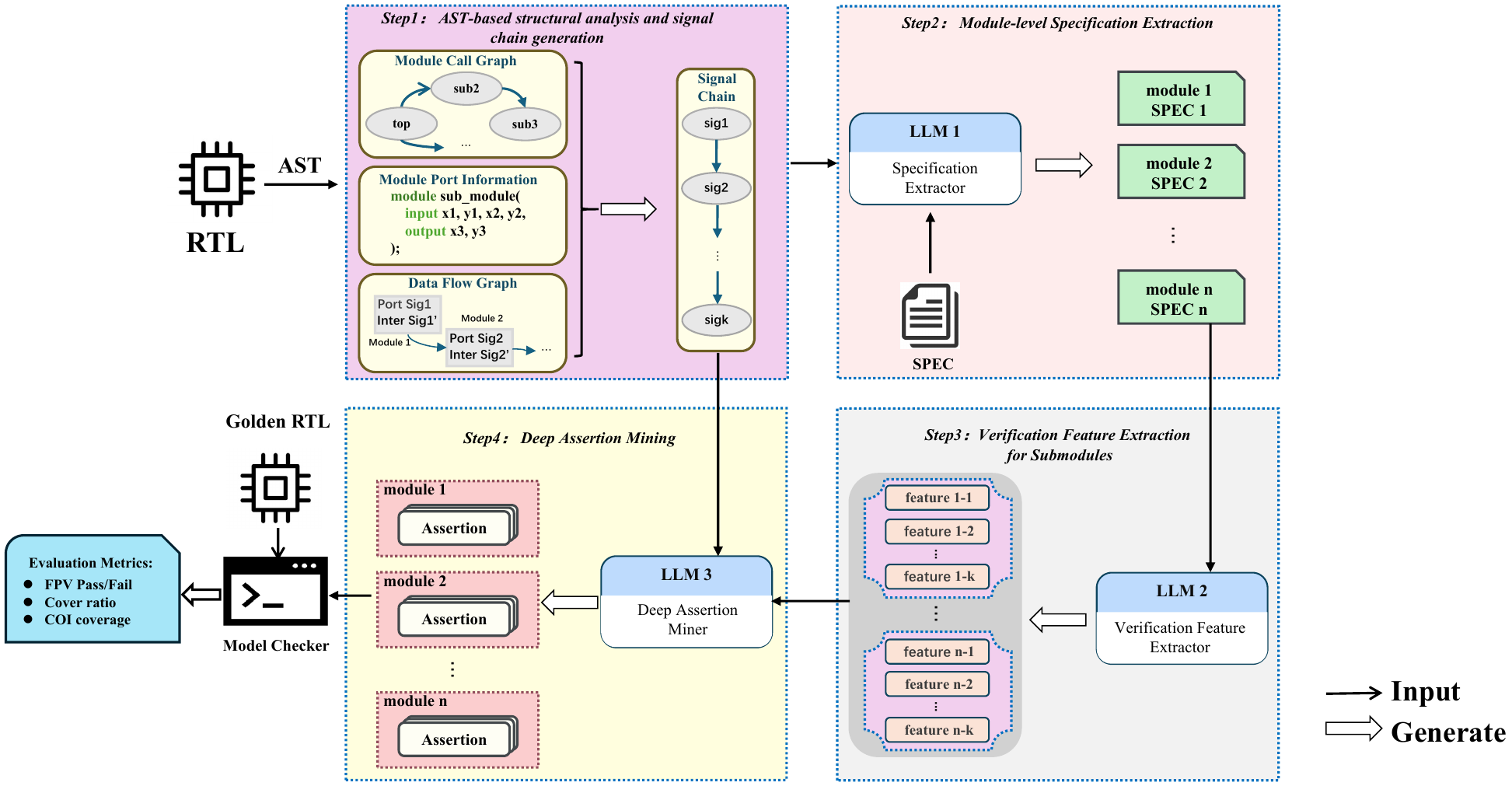}
\caption{The flowcharts of the AssertMiner framework for mining deep assertion, and the mined assertions are further evaluated against the golden RTL implementation using model checking tools.}
\label{fig:2}
\end{figure*}

Considering the complexity of generating assertions, Yan et al.\cite{b11} proposed AssertLLM, which decomposes the task of generating assertions from natural language specifications into multiple subtasks. On this basis, Wu et al.\cite{b16} proposed Spec2Assertion, which employs regularization and Chain-of-Thought prompting to generate high-quality assertions directly from specifications. However, their approach still struggled to bridge the semantic gap between specifications and RTL details. To address this, Bai et al.\cite{b5} proposed AssertionForge, a knowledge-graph-based framework, linking natural language specifications with RTL semantics to improve assertion quality. Since most approaches generate assertions only at the top-module level, Lyu et al.\cite{b27} proposed to achieve multi-layer assertion generation by constructing cross-layer signal bridging.

Overall, recent studies have begun to extend assertion generation to submodules. However, in design specifications, the descriptions of submodules are often vague and simplified, which hinders LLMs generating module-level assertions. In this paper, we’ll explore a dedicated framework for generating module-level spec, followed by mining effective deep assertions specifically targeting module level.

\section{The Framework of AssertMiner}
    
\subsection{Workflow Overview}
To address the challenge of generating assertions for modules reflecting micro-architectures of design, we propose the framework of AssertMiner. AST-based structural analysis is conducted to construct dedicated specifications for each module, enabling the mining of targeted deep assertions for module-level signals. Specifically, as illustrated in Fig. \ref{fig:2}, the entire framework is designed as a four-stage workflow for automatically generating deep assertions.

\subsection{Step 1: AST-based structural analysis and signal chain generation}

% In practical scenarios, the RTL code provided to verification engineers is assumed to accurately define the input and output ports for each module. Based on this assumption, we can directly extract the inter-module invocation relationship graph, port information, and signal propagation across module interfaces from the RTL code. This extraction allows us to understand the functionality of each module and the overall system behavior. To accurately extract the aforementioned information, our work introduces a dedicated LLM\textcircled{1} \texttt{Relationship Extractor} specifically for extracting the three parts from RTL code. We present the response generated by the LLM\textcircled{1} under guided prompting, as shown in Fig.\ref{fig:3} 

In practical scenarios, the RTL code provided to verification engineers is typically assumed to have a correct define of the input and output ports for each module. Based on this assumption, the module call graph (MCG), I/O table, the dataflow graph, and inter-module signal chains can be directly extracted from the RTL code. Such extraction helps in understanding both the functionality of individual modules and the overall system behavior. To accurately obtain the required static structure, we parse the RTL into an AST. The detailed process is described as follows:

\begin{enumerate}
    \item \textbf{Module Call Graph Extraction}: From AST instantiations, module hierarchy and interconnections are derived. An MCG is then constructed where nodes are modules and edges are instantiations.
    \item \textbf{I/O Table Extraction}: Again from the AST, we extract each module’s I/O declarations and resolve instantiation mappings to link ports to parent signals, producing a per-module I/O table (name, direction, connection), where direction refers to the port type which is either input or output, and connection refers to the connected ports of the upstream module or downstream module. 
    \item \textbf{Dataflow Graph Extraction}: Signal assignments and dependencies from the AST are analyzed to build a dataflow graph that represents how signals propagate within and across modules.
    \item \textbf{Signal Chain Extraction}:     \begin{figure}[h]
  \centering
  \includegraphics[width=1\linewidth,height=2.0\textheight,keepaspectratio]{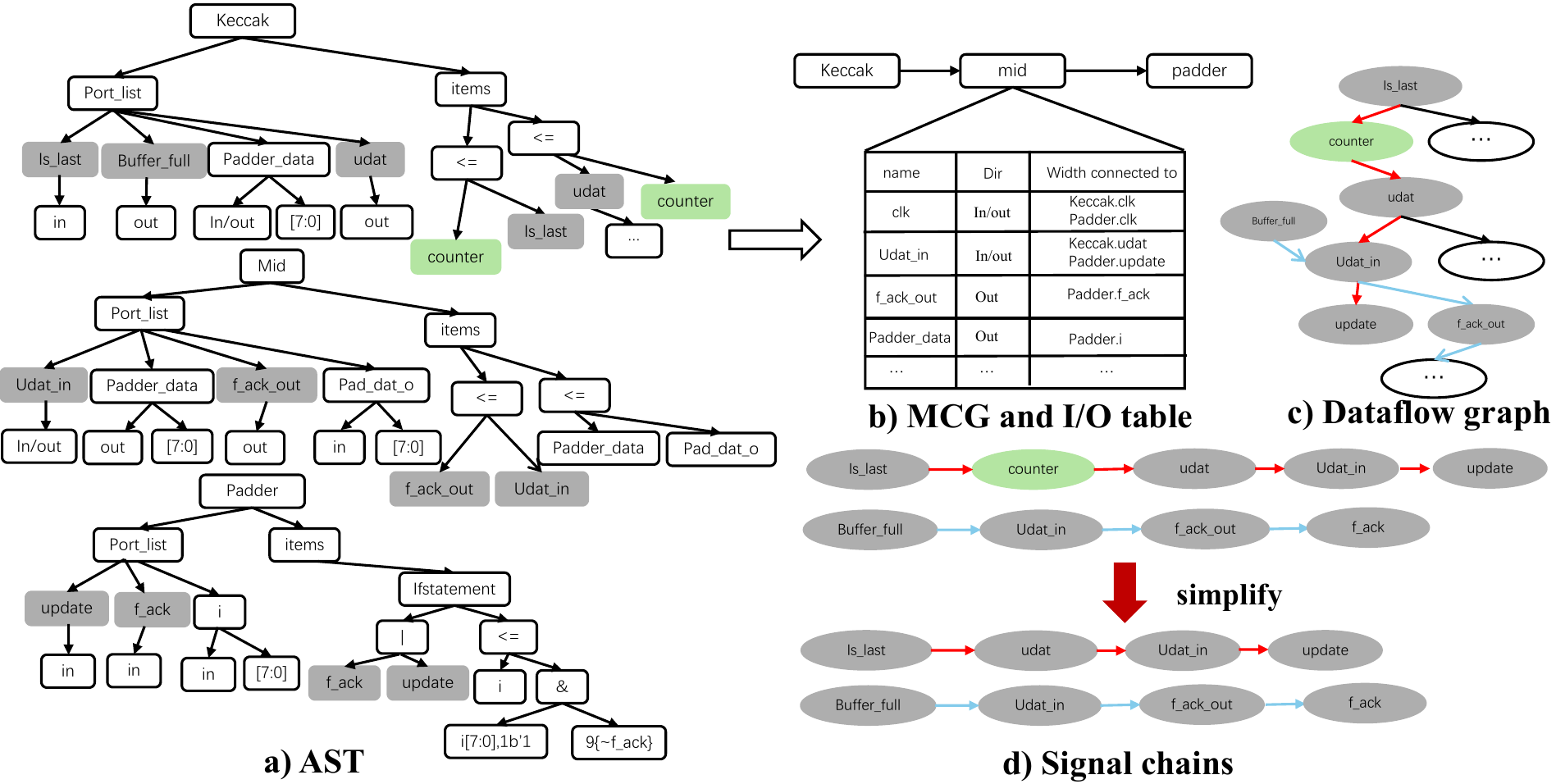}
  \caption{Static structural information extracted from the RTL code of SHA3}
  \label{fig:MCG}
\end{figure}AssertMiner leverages the MCG, I/O table, and data-flow graph to extract signal chains that trace inter-module signal propagation from top-level inputs to lower-level outputs. To guarantee robustness, internal module signals are pruned during chain simplification to prevent contamination from internal data flows which may include design bugs.
\end{enumerate}

For example, Fig. \ref{fig:MCG} presents the MCG, I/O table, data-flow graph, and signal chain results of SHA3 \cite{b28}, while only the signals and statements relevant to signal chains extraction are kept in the AST. Signal chains are generated through backward traversal of the data-flow graph, tracing each output to its input sources and forming concrete signal propagation paths that capture end-to-end dependencies within modules. After the concrete signal chains are generated, only the signals connected with I/O ports of modules are kept in the signal chains to get simplified signal chains, which are used in Step 2 for extracting reliable specifications for modules.

% \begin{figure}[h]
%   \centering
%   \includegraphics[width=0.9\linewidth]{fig/relationship.pdf}
%   \caption{Concise prompt and response of Relationship Extraction.}
%   \label{fig:3}
% \end{figure}
%先扔在这里，然后再进行处理

\begin{figure}[ht]
  \centering
  \includegraphics[width=0.9\linewidth,height=0.45\textheight,keepaspectratio]{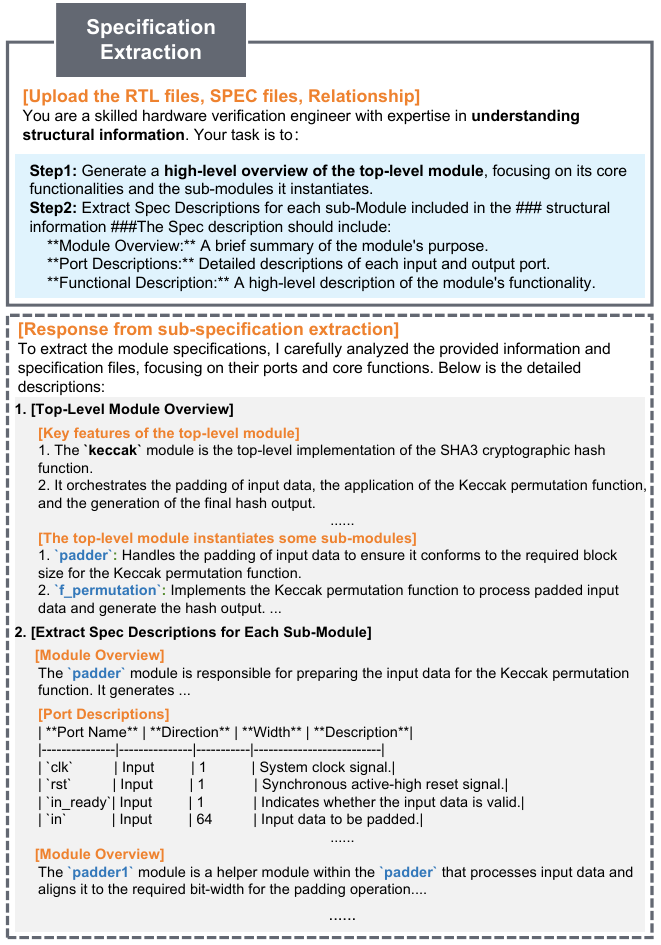}
  \caption{Concise prompt and response of module-level specification extraction.}
  \label{fig:4}
\end{figure}

\subsection{Step 2: Module-Level Specification Extraction}

Due to the lack of module-level specifications, designers need to interpret and supplement their functionalities based on personal understanding. This uncertainty makes it difficult to directly generate assertions related to the micro-architecture. Moreover, the RTL implementation itself may contain design errors, making it an unreliable source for specification inference.

A natural idea is that, if clear module-level specifications can be derived for each submodule, they can serve as a reliable foundation for generating micro-architecture-level assertions. In practical verification, module input and output port configurations are typically assumed to be correct. Therefore, by leveraging the structural information extracted in step 1, the overall functionality of each module can be inferred. Given the semantic understanding and contextual reasoning capabilities of LLMs, such inference is indeed feasible \cite{b29}.

\begin{figure}[ht]
  \centering
  \includegraphics[width=1\linewidth,height=1.3\textheight,keepaspectratio]{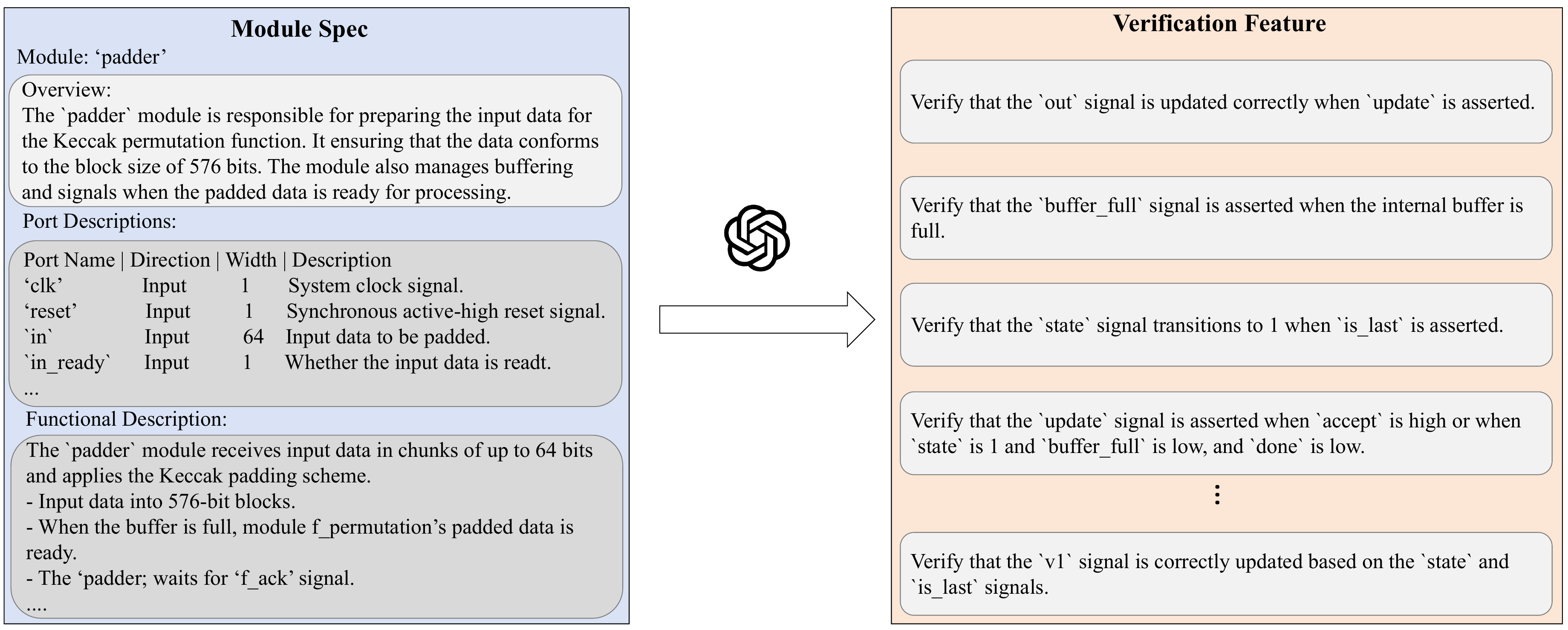}
  \caption{Example of decomposing moduel's spec into verification features}
  \label{fig:decompose}
\end{figure}
To achieve this goal, the proposed approach combines the overall functional intent described in the original design specification with the structural information  obtained in Step 1. An example of specification extraction is shown in Fig. \ref{fig:4}.

% To address the aforementioned challenge, we propose an LLM\textcircled{2} \texttt{Specification Extractor} that can accurately extract the specifications for modules without relying on the implementation details of the RTL code. Specifically, the LLM\textcircled{2} first understands the functional overview of the entire design provided in the original specification. Based on this understanding, the LLM\textcircled{2} further leverages the relationships extracted in Step 1 to accurately infer the overall functionality of a module and the behavior of each signal at its ports. Consequently, a detailed module-level specification is generated, as shown in Fig.\ref{fig:4}.
\subsection{Step 3: Verification Feature Extraction for Modules}

In practical top-down hardware development, it often lacks explicit alignment between the overall specification and submodule behaviors. To achieve fine-grained and traceable verification, verification engineers should decompose module-level specifications into fine-grained, verifiable features. Based on the specifications extracted in Step 2, we employ the LLM to identify verification features, i.e., concise atomic statements that directly specify what must be verified within each module. Using the ``padder'' module of SHA3 as an example, we generated the verification features as shown in Fig. \ref{fig:decompose}.

\subsection{Step 4: Deep Assertion Mining}

% Based on the information provided by the previous steps, we propose the LLM\textcircled{4} \texttt{Deep Assertion Generator} that can be guided to accurately generate deep assertions within modules. Furthermore, we have provided a standard template \texttt{\{source\_module\_name\}.\{signal\_name\}} to effectively constrain the generation behavior of LLM\textcircled{4}.

After obtaining verification features and signal chains from the previous steps, the next step is to automatically mine assertions that describe the signal behaviors within modules. Meanwhile, to ensure structural consistency among all mined assertions, a prompting strategy is applied to guide the generation with a unified output format, as shown below:
% \begin{equation*}
% \scriptsize
% \begin{aligned}
% &\texttt{assert property (@(posedge padder.clk) (padder.state == 1’b0} \\
% &\texttt{\&\& padder.in_ready \&\&!padder.buffer_full)} \\
% &\texttt{ |->padder.accept == 1’b1);} \\
% % &\quad \texttt{disable iff (!\{source\_module\_name\}.\{signal\_name\})}
% \end{aligned}
% \end{equation*}

\begin{lstlisting}[language=Verilog]
assert property (@(posedge <CLK>)
  (<AP> { && <AP> } { || <AP> } ...) |-> (<AP> { && <AP> } { || <AP> } ...)
);
\end{lstlisting}
where AP represents an atomic proposition, which can include multiple signals following the format \texttt{module.signal\_name}, referring to the signal location in the module hierarchy.

\section{Experiments}

\definecolor{lightblue}{RGB}{70,190,240}
\definecolor{lightgreen}{RGB}{152, 251, 152}

% \begin{table*}[h]
% \centering
% \caption{Performance of AssertMiner in generating sub-module deep assertions}
% \label{Performance comparison of AssertLLM, Spec2Assertion and AssertMiner}
% \begin{tabularx}{\textwidth}{
%   >{\centering\arraybackslash}X | 
%   *{4}{>{\centering\arraybackslash}X|} 
%   *{4}{>{\centering\arraybackslash}X|} 
%   *{3}{>{\centering\arraybackslash}X|} 
%   >{\centering\arraybackslash}X 
% }
% \toprule
% \multirow{2}{*}{Metric} & \multicolumn{4}{c|}{\cellcolor{gray!20}\textbf{AssertLLM}} & \multicolumn{4}{c|}{\cellcolor{gray!20}\textbf{Spec2Assertion}} & \multicolumn{4}{c}{\cellcolor{lightblue}\textbf{AssertMiner}} \\
% \cline{2-13}
% & I\textsuperscript{2}C & Pairing & ECG & SHA3 & I\textsuperscript{2}C & Pairing & ECG & SHA3 & I\textsuperscript{2}C & Pairing & ECG & SHA3 \\
% \midrule
% $N/S/P$ & 28/24/4 & 50/46/14 & 44/30/8 & 28/23/8 & 9/9/0 & 22/14/0 & 29/24/3 & 24/18/2 & 25/25/17 & 28/24/13 & 26/26/12 & 24/24/20 \\
% $NVR$(\%) & 33.33 & 100 & 36.67 & 45.45 & 100 & 100 & 100 & 64.29 & \cellcolor{lightgreen}100 & \cellcolor{lightgreen}100 & \cellcolor{lightgreen}92.86 & \cellcolor{lightgreen}100 \\
% $BFC$(\%) & 0 & 0 & 0 & 20 & 0 & 0 & 0 & 0 & \cellcolor{lightgreen}82.79 & \cellcolor{lightgreen}89.82 & \cellcolor{lightgreen}82.22 & \cellcolor{lightgreen}80 \\
% $SFC$(\%) & 0 & 0 & 0 & 31.71 & 0 & 0 & 0 & 0 & \cellcolor{lightgreen}83.06 & \cellcolor{lightgreen}88.66 & \cellcolor{lightgreen}80 & \cellcolor{lightgreen}82.93 \\
% $TFC$(\%) & 10.9 & 0 & 0 & 0.48 & 0 & 0 & 0.01 & 0.06 & 79.79 & 60.40 & 57.85 & 78.18 \\
% \bottomrule
% \end{tabularx}
% \end{table*}
    
\subsection{Experimental Setup}

The benchmark data used in this study are sourced from the open-source dataset in \cite{b28}, while each design in this benchmark includes a specification file and the corresponding golden RTL implementation code. We employ the commercial formal verification tool Cadence JasperGold (version: 21.12.002) for correctness analysis. All experiments are conducted on a server equipped with an Intel(R) Xeon(R) Gold 6148 CPU @ 2.40GHz.

To evaluate the effectiveness of AssertMiner, we conduct comparative experiments with two state-of-the-art (SOTA) assertion generation methods under consistent experimental settings:

\begin{itemize}
    \item \textbf{AssertLLM \cite{b11}:} This multi-agent method extracts signal descriptions from specifications, maps them to RTL signals, and generates pre-RTL assertions.
    \item \textbf{Spec2Assertion \cite{b16}:} This method employs LLMs with progressive regularization and Chain-of-Thought prompting to extract functional descriptions and directly generate pre-RTL assertions from design specifications.
\end{itemize}

All three methods utilize GPT-4o as the experimental model to produce the results.

\subsection{Evaluation Metrics and Benchmarks}

We utilize the metrics provided in JasperGold, while the detailed evaluation criteria is shown in Table \ref{Summary of Evaluation Metrics}. 

\begin{table}[h]
\centering
\renewcommand{\arraystretch}{1.4} % 调整行间距
\caption{Summary of Evaluation Metrics}
\label{Summary of Evaluation Metrics}
\begin{tabular}{c|c}
\hline
\hline
\cellcolor{gray!20}\textbf{Evaluation Metrics} & \cellcolor{gray!20}\textbf{Summary} \\ 
\hline
\makecell{$N$} & \makecell{\hspace*{-25pt} The number of generated SVAs}\\ 
\makecell{$S$} & \makecell{\hspace*{-10pt} The number of syntax-correct SVAs}\\
\makecell{$P$} & \makecell{\hspace*{-18pt} The number of FPV-passed SVAs}\\
\makecell{$NVR$} & \makecell{\hspace*{-18pt} Rate when SVA antecedent is true}\\
\makecell{\textbf{$BFC$}} & \makecell{\hspace*{-14pt} Branch coverage in formal analysis} \\ 
\makecell{\textbf{$SFC$}} & \makecell{\hspace*{-5pt} Statement coverage in formal analysis} \\ 
\makecell{\textbf{$TFC$}} & \makecell{\hspace*{-15pt} Toggle coverage in formal analysis} \\
\hline
\hline
\end{tabular}
\\
\scriptsize % 更小的字体
\raggedright % 设置文本左对齐
\vspace{+0.1cm}
*Note: The coverage metrics \textbf{$TFC$} are computed within the \textbf{Cones of Influence (COI)} covered by the mined assertions, where a larger \textbf{COI} encompasses more signals. Therefore, in each design experiment, we use the number of signals involved in the largest \textbf{COI} as the standard denominator for the calculation.
\end{table}

In the experiment, we extract the correct assertions that pass FPV validation and use JasperGold once again to calculate the relevant coverage metrics with the correct assertions. Four representative designs, I\textsuperscript{2}C, ECG, Pairing, and SHA3, are used to intuitively evaluate the capability of AssertMiner. The detailed descriptions of these four designs are presented in Table \ref{Summary of Design}, where LOC stands for number of lines of code.

% \begin{table}[h]
% \centering
% \renewcommand{\arraystretch}{1.4} % 调整行间距
% \caption{Summary of Designs}
% \label{Summary of Design}
% \setlength{\tabcolsep}{3.5pt}
% \begin{tabular}{c|c|c|c}
% \hline
% \hline
% \cellcolor{gray!20}\textbf{Design Name} & \cellcolor{gray!20}\textbf{Func. Description} & \cellcolor{gray!20}\textbf{LoC} &
% \cellcolor{gray!20}\textbf{Num. of Cells} \\ \hline
% \makecell{\texttt{I\textsuperscript{2}C}} & \makecell{\hspace*{-4pt} Serial communication protocol.} & \makecell{5369} & \makecell{756}\\
% \makecell{\texttt{SHA3}} & \makecell{\hspace*{-17pt} Hash function computation.} & \makecell{141185} & \makecell{22228}\\ 
% \makecell{\texttt{ECG}} & \makecell{\hspace*{-12pt} Biological signal acquisition.} & \makecell{398686} & \makecell{59084}\\ 
% \makecell{\texttt{Pairing}} & \makecell{\hspace*{-13pt} Cryptographic key exchange.} & \makecell{1561498} & \makecell{228287}\\ 
% \hline
% \hline
% \end{tabular}
% \end{table}

\begin{table}[h]
\centering
\caption{Summary of Designs}
\label{Summary of Design}
\begin{tabular}{|c|c|c|c|}
\hline
\rowcolor{gray!20}
\textbf{Design Name} & \textbf{Func. Description}&\textbf{LOC}& \textbf{Cell Num.}\\ \hline
\makecell{\textbf{I\textsuperscript{2}C}}  & \makecell{\hspace*{-4pt} Serial communication protocol.}& \makecell{1282}& \makecell{756}\\ \hline
\makecell{\textbf{ECG}}  & \makecell{\hspace*{-12pt} Biological signal acquisition.}& \makecell{1635}& \makecell{59084}\\ \hline
\makecell{\textbf{Pairing}}  & \makecell{\hspace*{-13pt} Cryptographic key exchange.}& \makecell{2145}& \makecell{228287}\\ \hline
\makecell{\textbf{SHA3}}  &  \makecell{\hspace*{-17pt} Hash function computation.}& \makecell{618}& \makecell{22228} \\ \hline
\end{tabular}
\end{table}

\subsection{Experimental Results}
\subsubsection{\textbf{Module-level Spec Generation Stability}}
\begin{figure}[H]
\centering
\includegraphics[width=1\linewidth]{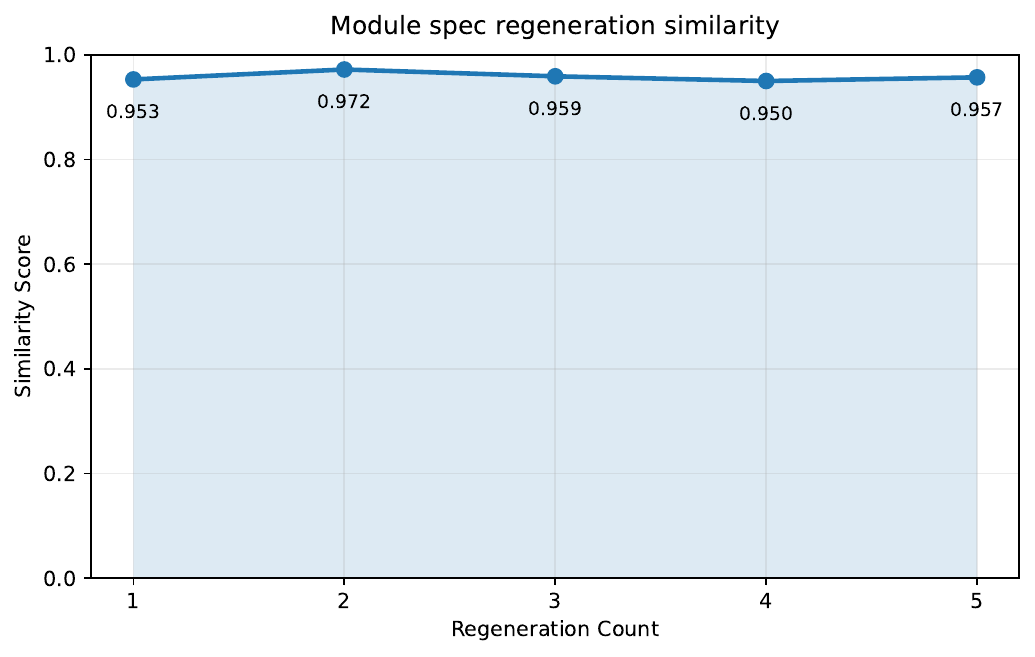}
\caption{Consistence of module spec regeneration}
\label{fig:Module spec regeneration similarity}
\end{figure}

To evaluate the stability of module-level specification generation, we repeatedly generated the specification of the I\textsuperscript{2}C module five times under identical input conditions. As shown in Fig. \ref{fig:Module spec regeneration similarity}, The results demonstrate that the proposed framework maintains stable performance in module-level specification generation. Despite the inherent randomness of LLMs, the regenerated specifications remain highly consistent, with an average similarity above 95\%.
\subsubsection{\textbf{Deep Assertion Mining by AssertMiner}}

Since both AssertLLM and Spec2Assertion generate assertions from specifications, without module-level specifications, they cannot directly generate deep assertions for modules.

% \begin{table}[h]
% \centering
% \caption{Performance of AssertMiner in generating deep assertions}
% \label{Performance of AssertMiner}
% \begin{tabular}{c|cccc}
% \toprule
% \rowcolor{gray!20} % 添加表头背景颜色
% \textbf{Metric} & \textbf{I\textsuperscript{2}C} & \textbf{Pairing} & \textbf{ECG} & \textbf{SHA3} \\
% \midrule
% $N/S/P$ & 25/25/17 & 28/24/13 & 26/26/12 & 24/24/20 \\
% $NVR$(\%) & \cellcolor{lightgreen}100 & \cellcolor{lightgreen}100 & \cellcolor{lightgreen}92.86 & \cellcolor{lightgreen}100 \\
% $BFC$(\%) & \cellcolor{lightgreen}82.79 & \cellcolor{lightgreen}89.82 & \cellcolor{lightgreen}82.22 & \cellcolor{lightgreen}80 \\
% $SFC$(\%) & \cellcolor{lightgreen}83.06 & \cellcolor{lightgreen}88.66 & \cellcolor{lightgreen}80 & \cellcolor{lightgreen}82.93 \\
% $TFC$(\%) & 79.79 & 60.40 & 57.85 & 78.18 \\
% \bottomrule
% \end{tabular}
% \end{table}

\begin{table}[h]
\centering
\caption{Performance of AssertMiner in Generating Deep Assertions}
\label{Performance of AssertMiner}

\renewcommand{\arraystretch}{1.2} % 调整行高
\setlength{\tabcolsep}{6.5pt}       % 调整列间距

\begin{tabular}{
    >{\raggedright\arraybackslash}p{1.3cm}  % 第一列左对齐，宽度固定
    *{4}{>{\centering\arraybackslash}p{1.3cm}}  % 后四列居中
}
\toprule
\rowcolor{headergray}
\textbf{Metric} & \textbf{I\textsuperscript{2}C} & \textbf{Pairing} & \textbf{ECG} & \textbf{SHA3} \\
\midrule
$N/S/P$        & 25/25/17 & 28/24/13 & 26/26/12 & 24/24/20 \\
\rowcolor{lightgreen!60} 
$NVR$(\%)      & 100 & 100 & 92.86 & 100 \\
\rowcolor{lightgreen!40} 
$BFC$(\%)      & 82.79 & 89.82 & 82.22 & 80 \\
\rowcolor{lightgreen!40} 
$SFC$(\%)      & 83.06 & 88.66 & 80 & 82.93 \\
$TFC$(\%)      & 79.79 & 60.40 & 57.85 & 78.18 \\
\bottomrule
\end{tabular}
\end{table}

Deep assertion mining, though highly challenging, is enabled by the proposed framework. As shown in Table \ref{Performance of AssertMiner}, although we do not exploit RTL internal implementation details inside modules, resulting in 17\% $\sim$ 54\% error rates $(P/N)$ of the generated assertions, AssertMiner achieves over 80\% coverage in terms of BFC \& SFC with only a small set of generated correct deep assertions.

\subsubsection{\textbf{Integration Experiments of AssertMiner with Previous Methods}}

% \begin{table*}[h]
% \centering
% \caption{Performance comparison of AssertLLM and AssertLLM + AssertMiner}
% \label{Performance comparison of AssertLLM and AssertLLM + AssertMiner}
% \begin{tabularx}{\textwidth}{
%   >{\centering\arraybackslash}X | 
%   *{4}{>{\centering\arraybackslash}X|} 
%   *{4}{>{\centering\arraybackslash}X} 
% }
% \toprule
% \multirow{2}{*}{Metric} & \multicolumn{4}{c|}{\cellcolor{gray!20}\textbf{AssertLLM}} & \multicolumn{4}{c}{\cellcolor{lightblue}\textbf{AssertLLM + AssertMiner}} \\
% \cline{2-9}
% & I\textsuperscript{2}C & Pairing & ECG & SHA3 & I\textsuperscript{2}C & Pairing & ECG & SHA3 \\
% \midrule
% $N/S/P$ & 127/112/58 & 32/32/12 & 44/44/19 & 31/31/26 & 171/164/74 & 60/56/25 & 70/70/31 & 55/55/46 \\
% $NVR$(\%) & 100 & 13.64 & 76.67 & 80 & 100 & \cellcolor{lightgreen}62 & \cellcolor{lightgreen}83.5 & \cellcolor{lightgreen}85 \\
% $BFC$(\%) & 80.23 & 76.12 & 82.22 & 92 & \cellcolor{lightgreen}82.84 & \cellcolor{lightgreen}82.82 & 82.22 & \cellcolor{lightgreen}100 \\
% $SFC$(\%) & 82.26 & 83.63 & 80.74 & 90.24 & \cellcolor{lightgreen}83.87 & \cellcolor{lightgreen}83.66 & 80.74 & \cellcolor{lightgreen}95.12 \\
% $TFC$(\%) & 78.93 & 75.44 & 60.8 & 89.41 & \cellcolor{lightgreen}80.61 & 75.44 & \cellcolor{lightgreen}62.29 & 89.41 \\
% \bottomrule
% \end{tabularx}
% \end{table*}
\begin{table*}[h!]
\centering
\caption{Performance Comparison of AssertLLM and AssertLLM + AssertMiner}
\label{Performance comparison of AssertLLM and AssertLLM + AssertMiner}

\setlength{\tabcolsep}{7pt}

\begin{tabularx}{\textwidth}{
  >{\raggedright\arraybackslash}p{2.1cm} |
  *{4}{>{\centering\arraybackslash}X|} 
  *{4}{>{\centering\arraybackslash}X|} 
  >{\centering\arraybackslash}p{1.8cm}
}
\toprule
\multirow{2}{*}{\textbf{Metric}} 
& \multicolumn{4}{c|}{\cellcolor{headergray}\textbf{AssertLLM}} 
& \multicolumn{4}{c|}{\cellcolor{headerblue}\textbf{AssertLLM + AssertMiner}} 
& \multirow{2}{*}{\textbf{Average Imp.}} \\
\cline{2-9}
& I\textsuperscript{2}C & Pairing & ECG & SHA3 
& I\textsuperscript{2}C & Pairing & ECG & SHA3 & \\
\midrule
$N/S/P$ & 127/112/58 & 32/32/12 & 44/44/19 & 31/31/26 
& 171/164/74 & 60/56/25 & 70/70/31 & 55/55/46 & -- \\
$NVR$(\%) & 100 & 13.64 & 76.67 & 80 
& 100 & 62 & 83.5 & 85 & \textbf{+15.55} \\
\midrule
$BFC$(\%) & 80.23 & 76.12 & 82.22 & 92 
& \cellcolor{lightgreen!50}82.84 & \cellcolor{lightgreen!50}82.82 & 82.22 & \cellcolor{lightgreen!50}100 
& \textbf{+4.33} \\
$SFC$(\%) & 82.26 & 83.63 & 80.74 & 90.24 
& \cellcolor{lightgreen!50}83.87 & \cellcolor{lightgreen!50}83.66 & 80.74 & \cellcolor{lightgreen!50}95.12 
& \textbf{+2.14} \\
$TFC$(\%) & 78.93 & 75.44 & 60.8 & 89.41 
& \cellcolor{lightgreen!50}80.61 & 75.44 & \cellcolor{lightgreen!50}62.29 & 89.41 
& \textbf{+1.39} \\
\bottomrule
\end{tabularx}
\end{table*}

% \begin{table*}[h]
% \centering
% \caption{Performance comparison of Spec2Assertion and Spec2Assertion + AssertMiner}
% \label{Performance comparison of Spec2Assertion and Spec2Assertion + AssertMiner}
% \begin{tabularx}{\textwidth}{
%   >{\centering\arraybackslash}X | 
%   *{4}{>{\centering\arraybackslash}X|} 
%   *{4}{>{\centering\arraybackslash}X} 
% }
% \toprule
% \multirow{2}{*}{Metric} & \multicolumn{4}{c|}{\cellcolor{gray!20}\textbf{Spec2Assertion}} & \multicolumn{4}{c}{\cellcolor{lightblue}\textbf{Spec2Assertion + AssertMiner}} \\
% \cline{2-9}
% & I\textsuperscript{2}C & Pairing & ECG & SHA3 & I\textsuperscript{2}C & Pairing & ECG & SHA3 \\
% \midrule
% $N/S/P$ & 90/89/49 & 45/41/15 & 35/29/19 & 42/42/28 & 115/114/66 & 73/65/28 & 61/55/31 & 66/66/48 \\
% $NVR$(\%) & 100 & 100 & 100 & 92.24 & 100 & 100 & 100 & \cellcolor{lightgreen}97.83 \\
% $BFC$(\%) & 87.87 & 83.58 & 79.51 & 90.89 & \cellcolor{lightgreen}89.51 & \cellcolor{lightgreen}83.82 & \cellcolor{lightgreen}83.12 & \cellcolor{lightgreen}96 \\
% $SFC$(\%) & 89.44 & 66.67 & 81.05 & 87.78 & \cellcolor{lightgreen}91.05 & \cellcolor{lightgreen}80.66 & 81.05 & \cellcolor{lightgreen}92.68 \\
% $TFC$(\%) & 87.85 & 51.04 & 78.55 & 78.31 & \cellcolor{lightgreen}88.55 & \cellcolor{lightgreen}67.19 & 78.55 & \cellcolor{lightgreen}78.36 \\
% \bottomrule
% \end{tabularx}
% \end{table*}

% {
% % ——行距更紧——
% \renewcommand{\arraystretch}{1.05}

% % ——可选：配合 booktabs 压缩横线留白——
% \setlength{\aboverulesep}{0pt}
% \setlength{\belowrulesep}{0pt}
% \setlength{\extrarowheight}{0pt}

\begin{table*}[h!]
\centering
\caption{Performance comparison of Spec2Assertion and Spec2Assertion + AssertMiner}
\label{Performance comparison of Spec2Assertion and Spec2Assertion + AssertMiner}

\setlength{\tabcolsep}{7pt}

\begin{tabularx}{\textwidth}{
  >{\raggedright\arraybackslash}p{2.1cm} |
  *{4}{>{\centering\arraybackslash}X|} 
  *{4}{>{\centering\arraybackslash}X|} 
  >{\centering\arraybackslash}p{1.8cm}
}
\toprule
\multirow{2}{*}{\textbf{Metric}} 
& \multicolumn{4}{c|}{\cellcolor{headergray}\textbf{Spec2Assertion}} 
& \multicolumn{4}{c|}{\cellcolor{headerblue}\textbf{Spec2Assertion + AssertMiner}} 
& \multirow{2}{*}{\textbf{Average Imp.}} \\
\cline{2-9}
& I\textsuperscript{2}C & Pairing & ECG & SHA3 
& I\textsuperscript{2}C & Pairing & ECG & SHA3 & \\
\midrule
$N/S/P$ & 90/89/49 & 45/41/15 & 35/29/19 & 42/42/28 
& 115/114/66 & 73/65/28 & 61/55/31 & 66/66/48 & -- \\
$NVR$(\%) & 100 & 100 & 100 & 92.24 
& 100 & 100 & 100 & 97.83 & \textbf{+1.40} \\
\midrule
$BFC$(\%) & 87.87 & 83.58 & 79.51 & 90.89 
& \cellcolor{lightgreen!50}89.51 & \cellcolor{lightgreen!50}83.82 & \cellcolor{lightgreen!50}83.12 & \cellcolor{lightgreen!50}96 
& \textbf{+2.65} \\
$SFC$(\%) & 89.44 & 66.67 & 81.05 & 87.78 
& \cellcolor{lightgreen!50}91.05 & \cellcolor{lightgreen!50}80.66 & 81.05 & \cellcolor{lightgreen!50}92.68 
& \textbf{+5.13} \\
$TFC$(\%) & 87.85 & 51.04 & 78.55 & 78.31 
& \cellcolor{lightgreen!50}88.55 & \cellcolor{lightgreen!50}67.19 & 78.55 & \cellcolor{lightgreen!50}78.36 
& \textbf{+4.23} \\
\bottomrule
\end{tabularx}
\end{table*}

To explore whether AssertMiner can boost the coverage of prior methods, we integrated it with AssertLLM and Spec2Assertion. Table \ref{Performance comparison of AssertLLM and AssertLLM + AssertMiner} and Table \ref{Performance comparison of Spec2Assertion and Spec2Assertion + AssertMiner} show the metric improvements before and after integration with a few deep assertions from AssertMiner. For instance, SFC and TFC are improved by 5.13\% and 4.23\% on average in comparison with the original SPEC2Assertion, demonstrating the enhanced capability to catch design bugs in corner cases.

\subsubsection{\textbf{Error Coverage Evaluation via Mutation Testing}} 

To evaluate the error coverage of generated assersions, we utilize the JasperGold tool to inject mutation errors into three designs and performe error coverage analysis.

\begin{figure}[h]
\centering
\includegraphics[width=1.01\linewidth]{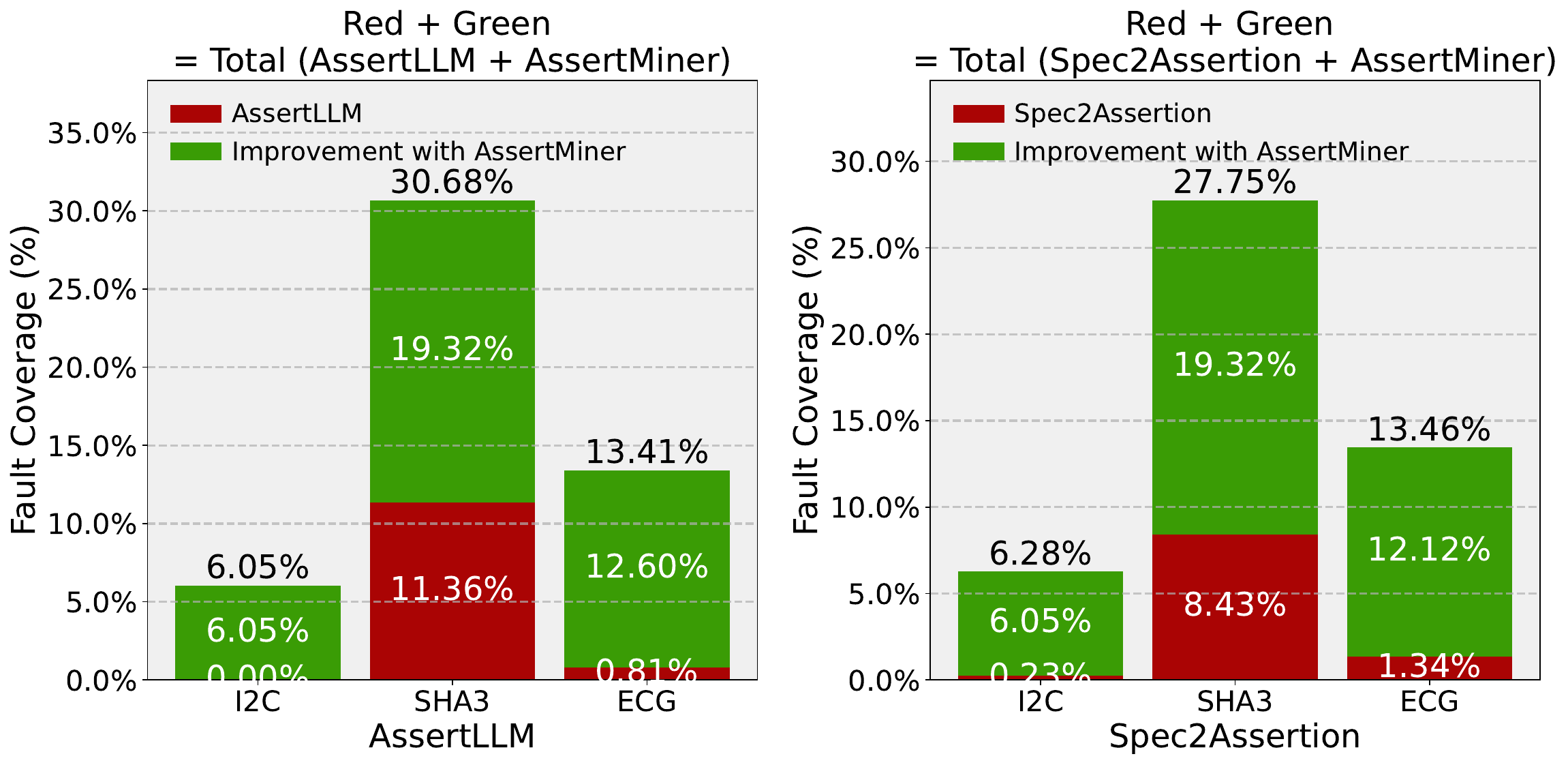}

\scriptsize % 更小的字体
\raggedright % 设置文本左对齐
*Note: Due to the large scale of the Pairing, it would require over 120 hours for mutation testing, so we exclude the Pairing from the experiment.
\caption{Error Coverage Enhancement in Mutation Testing via Integration of AssertLLM and Spec2Assertion with AssertMiner.}
\label{fig:9}
\end{figure}

\begin{figure}[h]
\centering
\includegraphics[width=0.95\linewidth]{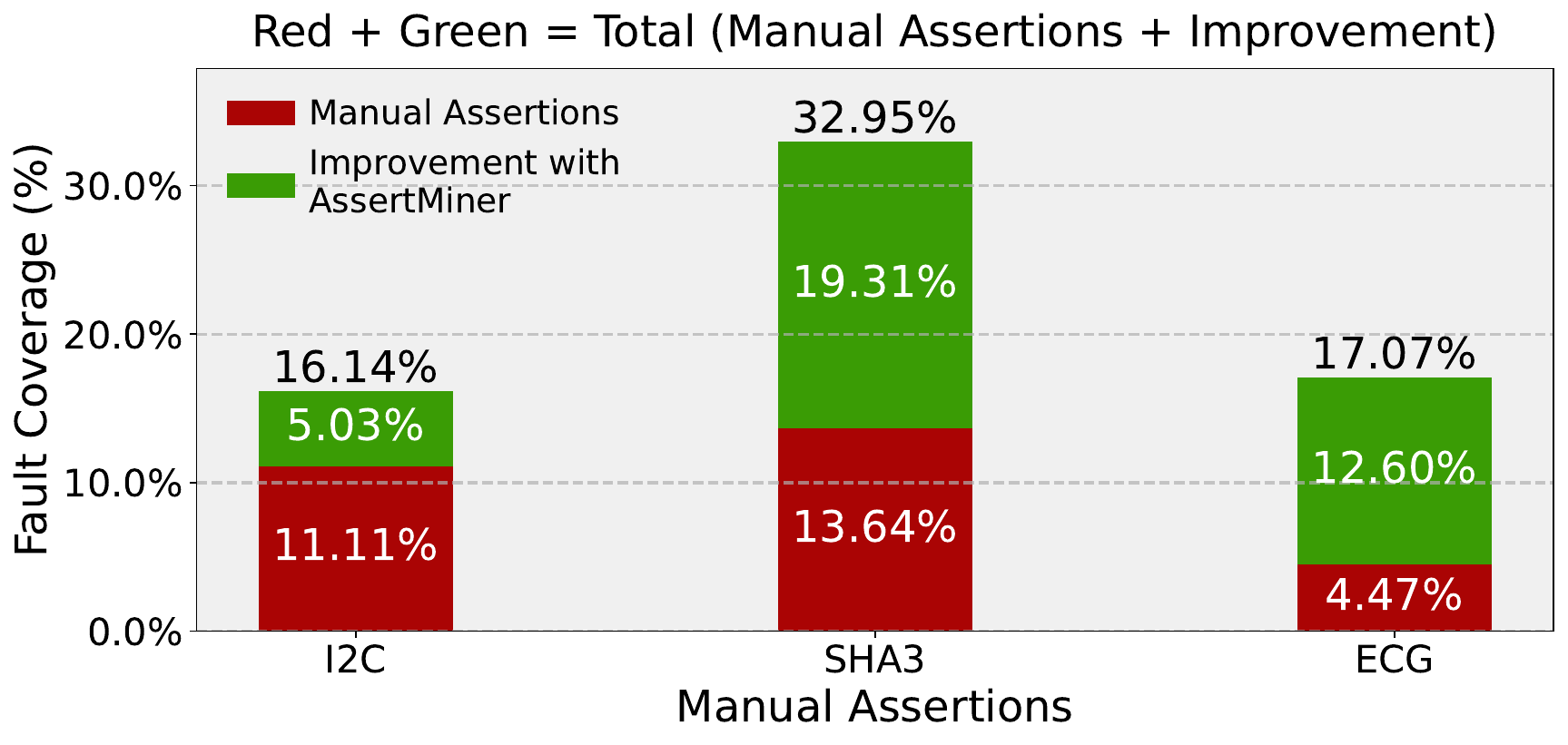}
\caption{Error Coverage Enhancement in Mutation Testing via Integration of Manually Designed Assertions with AssertMiner.}
\label{fig:12}
\end{figure}

This experiment aims to assess how well AssertMiner improves the effectiveness of existing verification methods. As Fig. \ref{fig:9} shows, integrating AssertMiner with the two prior methods boosts error coverage in mutation testing, highlighting the effectiveness of deep assertions. In contrast, assertions from the prior methods are often ineffective in catching errors, yielding low error coverage or even near 0\% for I2C.

To further assess the quality of deep assertions generated by AsserMiner, we also manually designed the same number of correct assertions using randomly selected key variables from the designs, and perform mutation testing. As Fig. \ref{fig:12} shows, AssertMiner's assertions still improve mutation error coverage by 5.0\% - 19.3\% even when combined with the high-quality manual assertions.

% Moreover, in our manual error injection attempts, we specifically activate an assertion in the top-level design and a deep assertion within a module, as shown in Fig. \ref{fig:10}. This setup allows us to compare the effectiveness of traditional top-level assertions with module-level deep assertions in detecting and localizing design errors.

% \begin{figure}[h]
% \centering
% \includegraphics[width=1\linewidth]{fig/mutation case study.pdf}
% \caption{A case of rapid assertion triggering and efficient debugging via assertion in the padder module of SHA3.}
% \label{fig:10}
% \end{figure}

% In this case study, injecting a single error into a module named \textit{padder} triggered both assertions simultaneously. The top-level assertion require ten clock cycles to trigger and could only infer an error's presence, while the module-level deep assertion trigger in just two clock cycles and directly identify the error signal, significantly reducing debugging time.

\section{Conclusion}
This paper presents AssertMiner, an automatic module-level assertion mining framework using multiple LLMs. By leveraging structural information extracted from the RTL code, AssertMiner infers module-level specifications and mines deep assertions. Experimental results across multiple design evaluations demonstrate that AssertMiner can be used to enhance the structural coverages of ABV, and more importantly, can achieve significant improvements of error detection capability over the existing SOTA methods. Future work will focus on improving the framework’s scalability to support more complex hierarchical designs and large-scale verification tasks.

\section*{Acknowledgment}

This paper is supported by the Chinese Academy of Sciences under grant No. XDB0660100 and No. XDB0660103. The corresponding authors are Zhiteng Chao and Huawei Li.

% \bibliographystyle{plain} % 使用 plain 样式
% % \bibliographystyle{IEEEtran}
% \bibliography{reference}

\end{document}